**Graphical Abstract**

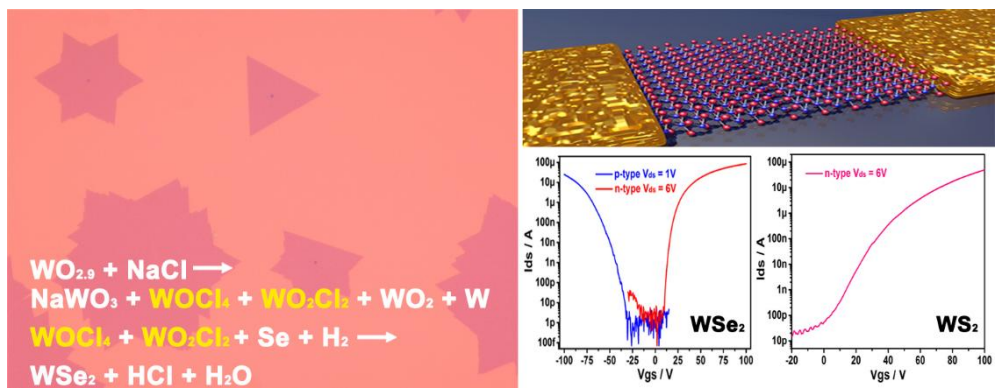

**Highlights:**

- We demonstrate chemical vapor deposition of $WSe_2$ and $WS_2$ monolayers, promoted by alkali metal halides, MX (M= Na or K; X=Cl, Br or I).

- Halide-stimulated the formation of volatile tungsten oxyhalides from tungsten oxide results in growing $WSe_2$ and $WS_2$ monolayers, in atmospheric pressure at a low and wide temperature window (700-850 °C).

- The as-grown $WSe_2$ and $WS_2$ monolayers show high hole and electron mobility up to 102 and 26 $cm^2 V^{-1} S^{-1}$ for $WSe_2$ devices and electron mobility of 14 for $WS_2$ devices.

# Halide-Assisted Atmospheric Pressure Growth of Large WSe$_2$ and WS$_2$ Monolayer Crystals

*Shisheng Li,* [1,2] *Shunfeng Wang,* [1,2] *Dai-Ming Tang,* [3] *Weijie Zhao,* [1,2] *Huilong Xu,* [1,2] *Leiqiang Chu,* [1,2] *Yoshio Bando,* [3] *Dmitri Golberg* [3] *and Goki Eda* *[1,2,4]

[1] Centre for Advanced 2D Materials, National University of Singapore, 6 Science Drive 2, 117546, Singapore

[2] Department of Physics, National University of Singapore, 2 Science Drive 3, 117542, Singapore

[3] World Premier International Center for Materials Nanoarchitectonics (MANA), National Institute for Materials Science (NIMS), Namiki 1-1, Tsukuba, Ibaraki, 305-0044, Japan

[4] Department of Chemistry, National University of Singapore, 3 Science Drive 3, 117543, Singapore

E-mail: g.eda@nus.edu.sg

**Abstract:** Chemical vapor deposition (CVD) of two-dimensional (2D) tungsten dichalcogenide crystals requires steady flow of tungsten source in the vapor phase. This often requires high temperature and low pressure due to the high sublimation point of tungsten oxide precursors. We demonstrate atmospheric pressure CVD of $WSe_2$ and $WS_2$ monolayers at moderate temperatures (700 ~ 850 ºC) using alkali metal halides (MX where M= Na or K and X=Cl, Br or I) as the growth promoters. We attribute the facilitated growth to the formation of volatile tungsten oxyhalide species during growth, which leads to efficient delivery of the precursor to the growth substrates. The monolayer crystals were found to be free of unintentional doping with alkali metal and halogen atoms. Good field-effect transistor (FET) performances with high current on/off ratio ~$10^7$, hole and electron mobilities up to 102 and 26 $cm^2$ $V^{-1}$ $s^{-1}$ for $WSe_2$ and electron mobility of ~14 $cm^2$ $V^{-1}$ $s^{-1}$ for $WS_2$ devices were achieved.

**Introduction**

Atomically thin crystals of transition metal dichalcogenides (TMDs) have recently emerged as a new class of two-dimensional (2D) materials and triggered intense research efforts [1,2]. Monolayer TMDs consist of a hexagonally packed layer of transition metal atoms sandwiched between two layers of chalcogen atoms. There are around 40 different types of TMDs with a wide variety of physical properties that make them attractive for fundamental studies as well as for practical applications [3,4]. Depending on the selection of the metal and chalcogen species, TMDs can be semiconducting (e.g. $MoS_2$, $WSe_2$), metallic (e.g. $NbS_2$, $TaSe_2$), or semimetallic (e.g. $WTe_2$) [3]. 2D TMD crystals exhibit unique electronic properties that are absent in the bulk materials due to geometrical confinement and distinct crystal symmetry [5-8].

Monolayers of group 6 TMDs have received significant attention for their direct bandgap in the visible range of wavelengths, high carrier mobility, excellent gate coupling, flexibility, and strong light-matter interaction [5,9-14]. These attributes make them appealing for ultrathin, lightweight, and flexible device applications [4,15,16].

Practical device applications of 2D TMDs rely on breakthroughs in the controlled and efficient growth of large-area films. While many groups have reported successful chemical vapor deposition (CVD) of various monolayer TMDs, the growth conditions and the quality of the film vary remarkably in the literature (Table S1 and S2, see also Ref [17]). One of the key challenges is the delivery of metal and chalcogen precursors to the growth substrates at a controlled flux. Typically, transition metal oxide and chalcogen in solid form are used as the precursors for the reaction. Since they have significantly different vapor pressures, the generation of steady vapor flux requires careful optimization of the temperature, pressure and carrier gas flow rate. Controlled growth of tungsten-based TMDs is particularly challenging due to the high sublimation temperature of tungsten oxide (e.g. $WO_3$ and $WO_{2.9}$). Recent reports on vapor transport and CVD growth of monolayer $WSe_2$ and $WS_2$ have shown that growth requires either high temperature (> 925 °C), or low pressure (~1 Torr) or both [13,14,18-24]. These conditions need to be met in order to achieve a sufficient flux of tungsten precursor in the vapor phase over a distance between the precursor to the growth substrates [14,19,24,25]. The high temperature and low pressure requirements are partly relaxed when oxide precursor is placed in direct contact with growth substrates [26,27].

However, this approach prevents controllable deposition of uniform large-area films. While the use of more volatile or gaseous tungsten-containing precursors such as $WCl_6$, $W(CO)_6$ and $WOCl_4$ has been demonstrated, the quality of the material or growth rate had to be compromised [28-31].

Halogen molecules such as $I_2$ and $Br_2$ are commonly used as the transport agents in chemical vapor transport (CVT) growth of bulk $WSe_2$ and $WS_2$ crystals thanks to their reactivity with metal oxide to form oxyhalide species such as $WO_2X_2$ and $WOX_4$ (X=Cl, Br or I), which have significantly lower melting point compared to $WO_3$ and $WO_{2.9}$ [32]. These transport agents are, however, not suitable for CVD growth of TMD monolayers due to their high reactivity and high vapor pressure.

In this paper, we report atmospheric pressure growth of high quality $WSe_2$ and $WS_2$ monolayers using a variety of alkali metal halides as the growth promoters. We show that presence of alkali metal halides allow the growth at temperatures as low as 700 ºC under atmospheric pressure. The facilitated growth suggests chemical reaction between the tungsten oxide precursor and the alkali metal halides and formation of volatile tungsten-based halide species. We found that the flakes grown by this method are highly crystalline, chemically pure, and exhibit good field-effect transistors (FETs) performances with hole and electron mobilities of 102 and 26 $cm^2\,V^{-1}\,s^{-1}$ for $WSe_2$ and electron mobility of ~14 $cm^2\,V^{-1}\,s^{-1}$ for $WS_2$ devices.

**Result and Discussion**

Figure 1a schematically illustrates the CVD setup used for growing tungsten based TMD monolayers. Tungsten oxide ($WO_{2.9}$) mixed with alkali metal halides (MX, M=Na, K; X= Cl, Br or I) and selenium/sulfur powders were employed as growth precursors. Alumina boat containing a mixture of 100 mg $WO_{2.9}$ and salt powders was loaded in the center of a 2-inch-diameter fused quartz tube. Growth substrates ($Si/SiO_2$) were placed ~8 mm above the $WO_{2.9}$ powders with its polished face down. Another boat containing chalcogen powder (~40 mg Se or ~50 mg S) was placed in the upstream region of the quartz tube.

As mentioned above, one of the main challenges in the growth of tungsten-based TMDs is the high sublimation temperature of tungsten oxides [19,21]. In the absence of alkali metal halides, no sublimation of tungsten oxide ($WO_{2.9}$) occurred under typical growth conditions (e.g. T = 850 °C, atmospheric pressure, flux of sulfur vapor in $Ar/H_2$ carrier gas) (Figures S1a and b). As a result, no deposition was observed on the growth substrate (Figure S1c). In stark contrast, $WO_{2.9}$ mixed with NaCl was found to undergo dramatic weight loss and color change after similar growth procedure, as can be seen from visual inspection of the crucible (Figures S1d and e). Deposition of large, triangular $WS_2$ crystals was observed on the growth substrate (Figure S1f).

Figures 1b-e show the optical microscope (OM) images of $WSe_2$ monolayers grown using KCl, KBr, KI and NaCl, respectively. Growth with KCl, KBr and NaCl at 825 °C resulted in deposition of quasi-hexagonal $WSe_2$ monolayers of various sizes. Large crystals had edge length of ~140 μm with a total area of ~100,000 μm². The high-magnification OM images of these polygonal $WSe_2$ monolayers reveal sharp straight

edges suggesting that they are highly crystalline (inset of Figures 1b, c and e). The nucleation of second layer was also observed in some monolayer flakes.

We found that the optimal growth temperature depends on the alkali metal halides used. For example, the optimal growth temperature of both $WSe_2$ and $WS_2$ using KCl, KBr and NaCl as the growth promoters was around 825 °C (Figures 1b, c, e, f and i), whereas the growth of monolayers occurred only below 800 °C when KI was used. In fact, the growth of $WSe_2$ and $WS_2$ monolayers was achieved at temperature as low as 700 °C as shown in the inset of Figures 1d and g. The optimal growth temperature roughly corresponded to the melting point of the alkali metal halides. For example, growth occurred at higher temperatures for NaCl, which has a melting point of $T_m^{NaCl}$ = 801 °C, and at lower temperatures with KI, which has the lowest melting point ($T_m^{KI}$ = 681 °C) among the four alkali metal halides used in this study. We found that the growth is most efficient in terms of yield and crystal quality when KCl, NaCl and KBr are used as the growth promoters and when the growth temperature is between 825 and 850 °C.

The low temperature atmospheric growth and halide-dependent growth temperatures point towards chemical reaction between $WO_{2.9}$ and the alkali metal halides that results in formation of volatile oxyhalide species. We conducted X-ray diffraction (XRD) analysis on the reaction product of $WO_{2.9}$ and NaCl after a typical CVD process at 850 °C without Se/S and found that it contains $NaWO_3$, $WO_2$ and W phases (Figure S2). We attribute our halide-assisted CVD growth of $WSe_2$ and $WS_2$ to the *in-situ* formation of $WO_2Cl_2$ ($T_m^{WO_2Cl_2}$ = 265 °C) and $WOCl_4$ ($T_m^{WOCl_4}$ = 211 °C) [32]. Note that $WOCl_4$ has been previously used as precursor for the CVD growth of WS$_2$ [31]. One possible reaction route in our system is

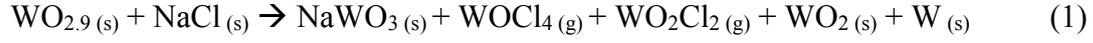

$$WO_{2.9 (s)} + NaCl_{(s)} \rightarrow NaWO_{3 (s)} + WOCl_{4 (g)} + WO_2Cl_{2 (g)} + WO_{2 (s)} + W_{(s)} \qquad (1)$$

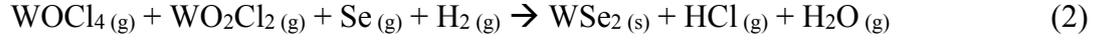

$$WOCl_{4 (g)} + WO_2Cl_{2 (g)} + Se_{(g)} + H_{2 (g)} \rightarrow WSe_{2 (s)} + HCl_{(g)} + H_2O_{(g)} \qquad (2)$$

The size, shape, quality, and yield of monolayer WSe$_2$ and WS$_2$ depended not only on the growth temperatures and the choice of alkali metal halides but also on other factors, such as the amount of chalcogen precursors. For example, the shape of WS$_2$ monolayers changed from triangles to hexagons and their average size increased substantially when the sulfur supply was reduced from ~50 to ~10 mg (Figures 1f and i). The quasi-hexagonal WS$_2$ crystals had edge lengths exceeding 200 μm with a total area over 200,000 μm$^2$, larger than the largest triangular crystals obtained by CVD techniques [23,24].

Below, we present a comprehensive characterization of triangular WSe$_2$ and WS$_2$ monolayers grown at 850 °C (Figures 1h and j). Figures 2a and d show OM images of a typical triangular WSe$_2$ and WS$_2$ monolayer with an edge length of ~50 μm. The AFM images of the monolayers (Fig 2b and e) show an edge step height ~1.2 nm, which is larger than the intrinsic height of ~ 0.7 nm, possibly due to surface adsorbates and roughness of the Si/SiO$_2$ surfaces. It is worth noting that the WSe$_2$ monolayers grown on flat sapphire surface show a thickness closer to 0.7 nm (Figure S3).

Photoluminescence (PL) mapping images of triangular WSe$_2$ and WS$_2$ monolayers (Figures 2c and f) reveal non-uniform emission intensity across the flakes. Stronger emission from the edges of WSe$_2$ and the center of WS$_2$ flakes suggests the presence of crystal defects [33]. Nevertheless, no internal grain structures are evident from the PL image, suggesting that these triangular flakes consist of a single crystalline domain [34]. The PL spectra in Figure 2g show narrow emission peaks centered at 748 and 620 nm for WSe$_2$ and WS$_2$ flakes respectively, in agreement with the reported direct optical band gap of these materials at room temperature [5]. The full width at half maximum (FWHM) values of ~50 meV and ~54 meV for WSe$_2$ and WS$_2$, respectively, are close to those of mechanically exfoliated monolayers and of CVD monolayers in previous reports [5,19,35]. Low temperature PL spectra show no evidence of trion and bound exciton peaks, further indicating that the concentration of defects is not significantly higher than that in typical mechanically exfoliated samples [20,36]. Figure 2h depicts representative unpolarized Raman spectra of WSe$_2$ and WS$_2$ monolayers with 532 nm and 473 nm excitations. CVD-grown WSe$_2$ monolayers exhibit two characteristic peaks at 250.5 cm$^{-1}$ and on 261.9 cm$^{-1}$. Similarly, WS$_2$ monolayers reveal two characteristic peaks located at 359.6 cm$^{-1}$ and 419.6 cm$^{-1}$, which are the well-known $E_{2g}^1$ and $A_{1g}$ modes. All Raman signatures are in a good agreement with those of mechanically exfoliated monolayer samples suggesting absence of significant strain and doping in our CVD samples [37-39].

Figures 3a and b show transmission electron microscope (TEM) images of a monolayer WSe$_2$ flake, revealing its highly crystalline nature. The lattice constant measured from the high-resolution TEM image was $a$ = 0.33 nm, consistent with that of 2H-WSe$_2$ [40]. The selected area electron diffraction (SAED) pattern further confirms the hexagonal crystal structure of the sample.

The chemical states and purity of the CVD-grown samples were studied by X-ray photoemission spectroscopy (XPS). Figures 3c and d show the binding energy of W *4f*, W *5p* and Se *3d* core levels. The spectra can be fitted with W *4f$_{7/2}$* (32.3 eV) and W *4f$_{5/2}$* (34.5 eV) doublets and a W *5p$_{3/2}$* peak (37.8 eV). The Se *3d* core levels can be fitted with Se *3d$_{5/2}$* (54.5 eV) and Se *3d$_{3/2}$* (55.4 eV) peaks in agreement with the spectra of pristine 2H-WSe$_2$ [41]. The atomic ratio between W and Se elements was 1:2.04, suggesting the absence of major Se vacancies in our samples. Similarly, the core level spectra of WS$_2$ were found to be consistent with those of bulk 2H-WS$_2$. The W *4f* and *5p* core levels show peaks at 33.2, 35.4, and 38.8 eV, corresponding to W *4f$_{7/2}$*, W *4f$_{5/2}$* and W *5p$_{3/2}$*, respectively (Figure 3e). The peaks corresponding to the S *2p$_{3/2}$* and S *2p$_{1/2}$* orbital of divalent sulfide ions are observed at 162.8 and 164 eV (Figure 3f) [42-44]. The blue shift (~0.9 eV) of W *4f* and *5p* peaks of WS$_2$ with respect to those of WSe$_2$ is attributed to the stronger W-S bonds compared to W-Se bonds. These results indicate that the monolayers are neither oxidized nor covalently functionalized with impurity species. XPS fine scan also showed no signatures of Na and Cl peaks, indicating that our samples are free of contamination due to the alkali metal halides (Figures S4a-d). This is also supported by our energy-dispersive X-ray spectroscopy (EDX) analysis of individual flakes in TEM (Figure S4e).

The electronic quality of the CVD-grown samples was studied by electrical measurements. The monolayers on the growth Si/SiO$_2$ substrates were electrically contacted using standard electron beam lithography and thermal evaporation. Palladium and silver were used for the realization of p-type and n-type WSe$_2$ devices, respectively. The large work function of Pd (5.2~5.6 eV), which is close to the valence band of WSe$_2$, allows efficient injection of holes into the WSe$_2$ channel [10,45]. On the other hand, lower work function of Ag (4.26eV) has been shown to be suitable for electron injection and observation of n-type behaviors.[12] In order to facilitate the p-type behavior of Pd-contact WSe$_2$ devices, the contacts were improved by exposing them to ozone prior to measurements [46]. For WS$_2$ devices, Au contacts were used for electron injection [47]. Figure S5 shows the schematic and corresponding OM images of the WSe$_2$ and WS$_2$ devices. Figures 4a and b show the transfer and output characteristics of typical p-type and n-type WSe$_2$ devices. These devices demonstrate a high current modulation by back-gate bias, with current on/off ratios of up to $10^7$. The linear output curves suggest that nearly Ohmic contacts are achieved for p-type and n-type devices. This is an indication that Fermi-level is not pinned to surface defect states of WSe$_2$, further supporting the high electronic quality of the sample. The WS$_2$ devices showed similarly high on/off ratios but predominantly n-type character despite the large work function of Au contacts (Figures 4c and d). The non-linear output curves at low V$_{ds}$ suggest the formation of Schottky barrier [24,47]. This observation may be explained by Fermi level pinning as commonly observed in MoS$_2$ devices [48].

The field-effect mobility of the devices was extracted from the linear region of the transfer curves using: $\mu = [dI_{ds}/dV_{gs}] \times [L/W]/[C \cdot V_{ds}]$, where $L$ is channel length, $W$ is channel width, and $C$ is the capacitance of the 300 nm-thick SiO$_2$. The extracted hole and electron mobilities of WSe$_2$ were found to be 102 and 26 cm$^2$ V$^{-1}$ s$^{-1}$, respectively. WS$_2$ devices showed electron mobility of 14 cm$^2$ V$^{-1}$ s$^{-1}$. These values are among the largest values reported for both exfoliated and CVD monolayers (Table S1 and S2) [10,14,19, 21,24,47,49,50].

**Conclusions**

In conclusion, we developed a halide-assisted atmospheric pressure growth of tungsten-based TMDs. We found that various alkali metal halides facilitate the transport of tungsten to the growth substrates by reacting with tungsten oxide to form volatile tungsten oxyhalide species. We showed that large, highly crystalline $WSe_2$ and $WS_2$ monolayers can be grown at temperatures as low as 700 °C at atmospheric pressure using this method. The high electronic quality of the samples was confirmed by the sharp photoluminescence and good field-effect transistor performance. Our halide-assisted CVD allows the growth of monolayers at moderate conditions and offers a flexible growth parameter window, which is advantageous for van der Waals hetero-epitaxy on reactive surfaces.

**Material and Methods**

*Synthesis of $WSe_2$ and $WS_2$ monolayers*: A 2-inch diameter horizontal fused quartz tube furnace was used. 1) For $WSe_2$ monolayer growth, ~40 mg Se (Alfa Aesar, 99.999+%, 1-5 mm) shot was placed in the upstream region of the furnace which reached ~450 °C during the growth. 100 mg mixture of $WO_{2.9}$ (Alfa Aesar, 99.99%, 150 mesh) /MX [M=Na or K; X=Cl, Br or I (GCE Laboratory chemicals, 99.95+%)] was loaded in the crucible at the center of the furnace. It is worth to note that $WO_{2.9}$ is better than $WO_3$ for growing $WSe_2$ monolayers (Figure S6). Therefore, $WO_{2.9}$ was used as tungsten precursor in this study. The optimal mass ratio of oxide and salt was between 9:1 to 17:3. The tube furnace was first pump down to ~1 Pa to remove air and moisture. Then, 80/20 sccm Ar/$H_2$ carrier gas was introduced until atmospheric pressure was achieved. The furnace was heated with a ramp rate of 35 °C/min to the growth temperatures (700 to 850 °C) and held at this temperature for 5 min before cooling down. 2) For the growth of WS$_2$, 10~50 mg S powder (GCE Laboratory chemicals, 99.95+%) was loaded in the upstream region, which reached a temperature of ~200 °C during the growth. Similiar to the WSe$_2$ growth, 100 mg mixture of WO$_{2.9}$/MX (MX=NaCl or KI) (GCE Laboratory chemicals, 99.95+%) was loaded in the crucible at the center of the furnace. The optimal mass ratio was 7:3.

*TEM sample preparation:* A drop of water was firstly dripped on the silicon substrate with grown WSe$_2$ monolayers. Then, the flakes were detached from substrate and floated on the water droplet. Copper grids with porous carbon film were used to scoop the flakes for TEM characterizations.

*Characterization of WSe$_2$ and WS$_2$ monolayers*: The morphology, microstructure, and chemical composition of the monolayers were studied using optical microscope (Olympus, BX51), atomic force microscope (Bruker Dimension FastScan, Tapping mode), transmission electron microscope (JEOL 3100FEF, 300 KeV), X-ray photo-electron spectroscope (Kratos Analytical, Axis Ultra DLD, Al Kα) and X-ray powder diffraction system (Philips X'Pert MPD). Raman and PL spectra were obtained with a laser confocal microscope (NT-MDT, NTEGR Spectra) in back scattering geometry with 532 and 473 nm excitation lasers.

*Device fabrication and test:* Standard e-beam lithography procedures were performed to contact the samples with metal electrodes. To achieve Ohmic contacts for both p- and n-type devices, different contact metals with were used. For p-type $WSe_2$ devices, a trilayer Cr/Pd/Au (1/10/40 nm) film was thermally evaporated. The p-type character of $WSe_2$ devices was enhanced by exposing it to ozone in an UV-ozone generator (details will be presented elsewhere). For n-type $WSe_2$ devices, trilayer Ag/Al/Au(5/15/40 nm) electrodes were used. For $WS_2$ devices, bilayer Cr/Au (0.5/50 nm) electrodes were used. All the tests were conducted in a glove box filled with nitrogen gas.

**Appendix A. Supplementary data**

Tables summarizing conditions for growing atomically-thin $WSe_2$ and $WS_2$ flakes. Comparison of growth with and without NaCl. Growth of $WSe_2$ monolayer on a sapphire substrate. XRD spectrum of $WO_{2.9}$/NaCl reaction products after a typical CVD growth procedure. XPS spectrum of Na *1s* and Cl *2p* for $WSe_2$ and $WS_2$ samples. EDX spectrum of $WSe_2$ samples. Schematics and OM images of $WSe_2$ and $WS_2$ devices. Supplementary data associated with this article can be found, in the online version, at doi:10.1016/j.apmt.2015.xx.xxx.

**Acknowledgements**

This research is supported by the National Research Foundation, Prime Ministers Office, Singapore under its Medium-sized Centre Programme as well as the grant NRF-NRFF2011-02 (G.E.). S.L. would like to thank Dr. J. Wu for his helpful discussion.

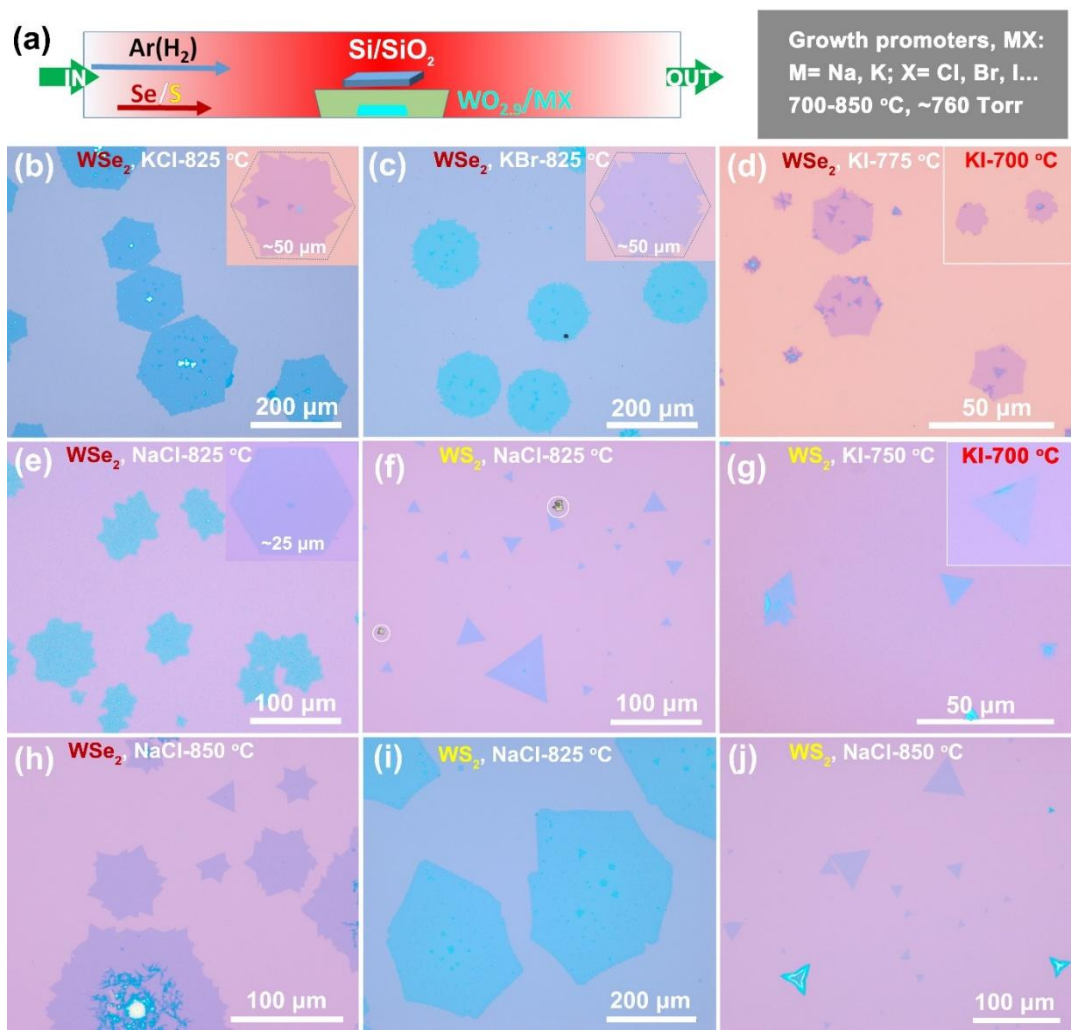

**Figure 1.** (a) Schematic illustration of the growth set up. $WO_{2.9}$ mixed with alkali metal halides and Se/S powders are loaded in the centre and upstream region of the CVD tube furnace. (b-e) OM images of the $WSe_2$ monolayers grown with (b) KCl, (c) KBr, (d) KI, (e) NaCl as the growth promoters at 825 °C. Insets of (b, c and e) are high-magnification OM images of polygon-shaped crystals with hexagonal symmetry. Inset of (d) is an OM image of $WSe_2$ monolayers grown at 700 °C. (f) An OM image of triangle $WS_2$ monolayers grown with NaCl as the growth promoter at 825 °C with ~50 mg S. (g) An OM image of triangular $WS_2$ monolayers grown by KI-assisted CVD at 750 °C. Inset is an OM image of $WS_2$ grown at 700 °C. (h) An OM image of $WSe_2$ monolayers grown at 850 °C. (i) An OM image of hexagonal $WS_2$ monolayers grown at 825 °C with ~10 mg S. (j) An OM image of $WS_2$ monolayers at 850 °C.

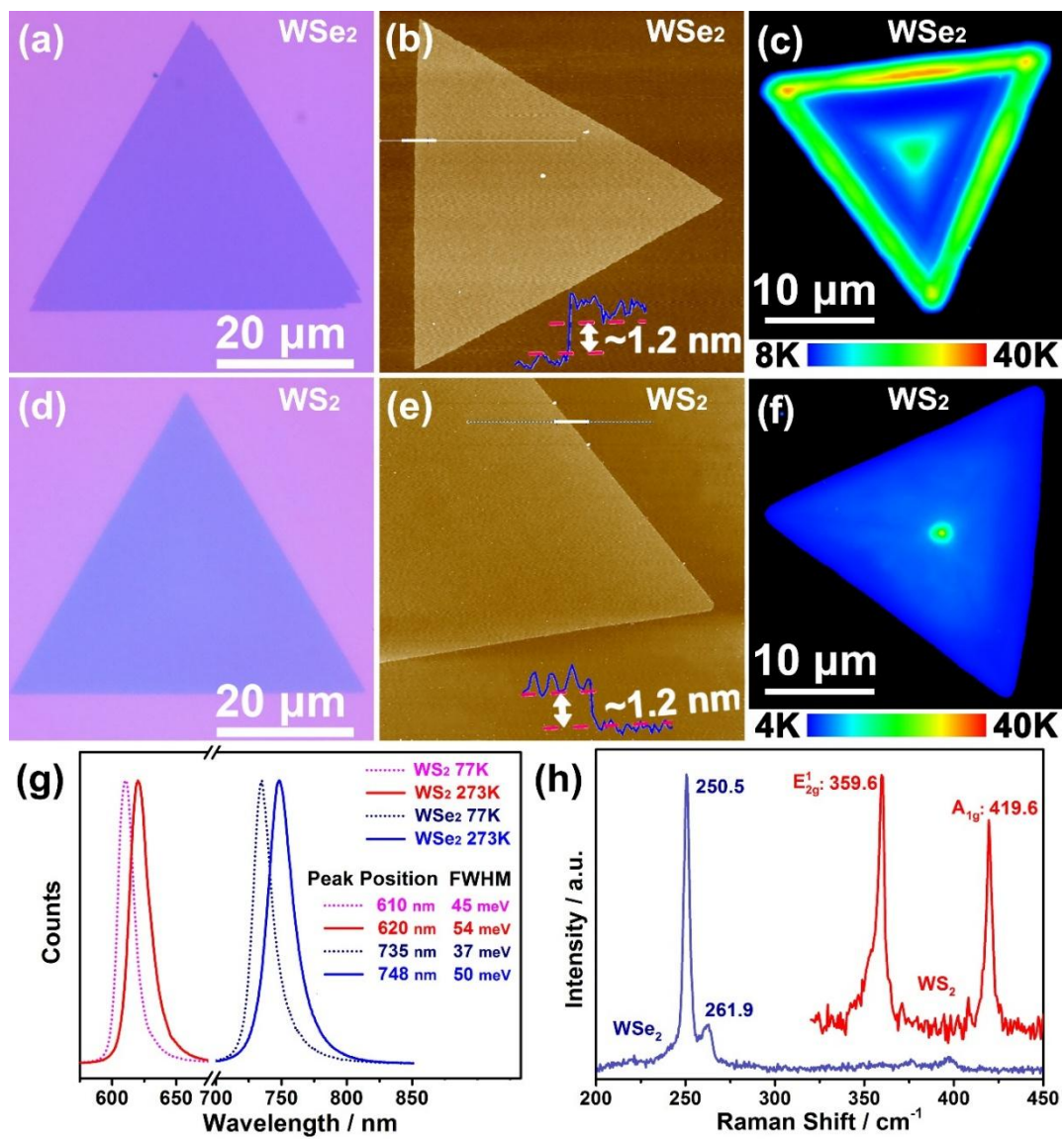

**Figure 2.** (a) OM, (b) AFM and (c) PL intensity mapping images of a triangular WSe$_2$ monolayer grown at 850 °C. (d) OM, (e) AFM and (f) PL intensity mapping images of a triangular WS$_2$ monolayer grown at 850 °C. (g) PL spectra of WSe$_2$ monolayer (blue) and WS$_2$ monolayer (red) at 77 K (dashed lines) and 273 K (solid lines). (h) Raman spectra of WSe$_2$ monolayer (blue) and WS$_2$ monolayer (red).

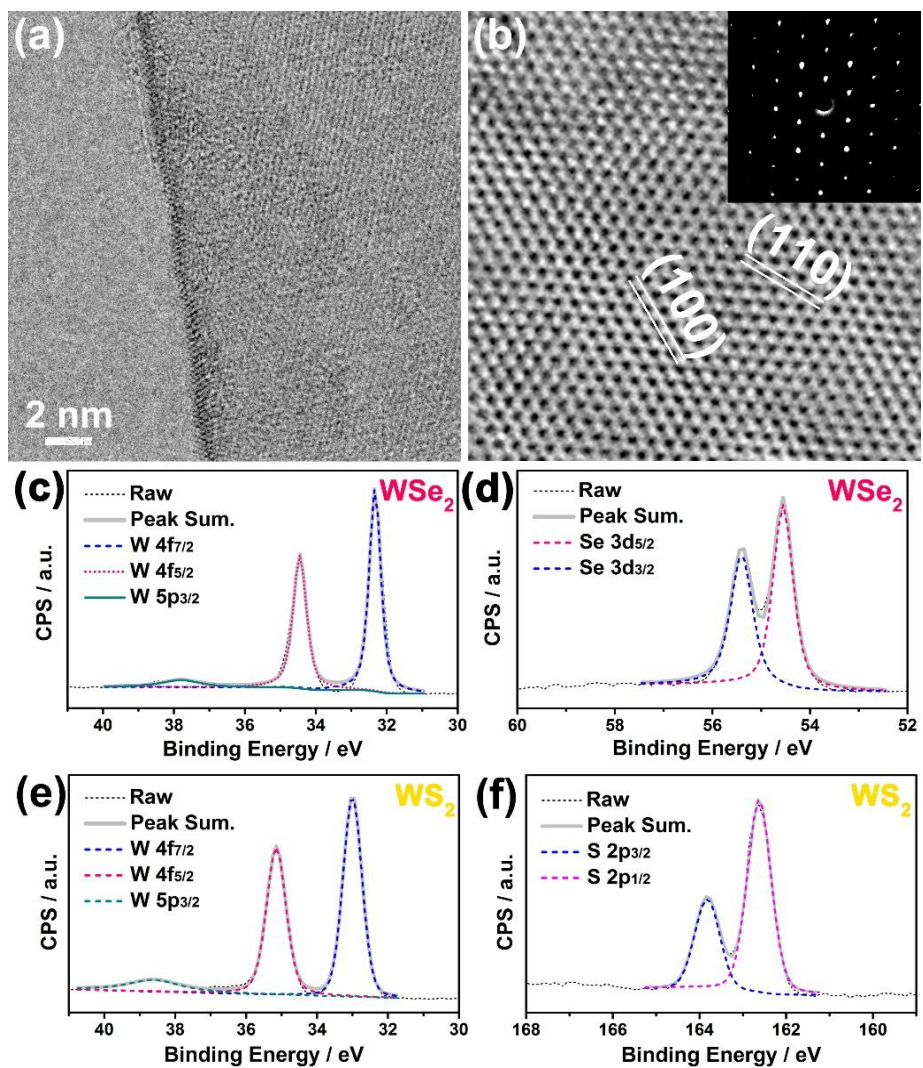

**Figure 3.** (a) Lower magnification and (b) high resolution (frame size: 8 x 8 nm$^2$) TEM image of a WSe$_2$ monolayer. Inset of (b) shows an SAED pattern. (c, d) XPS spectra showing (c) W *4f* and W *5p*, (d) Se *3d* regions of as-grown WSe$_2$ monolayers. (e, f) XPS spectra showing (e) W *4f* and W *5p*, and (f) S *2p* regions of as-grown WS$_2$ monolayers.

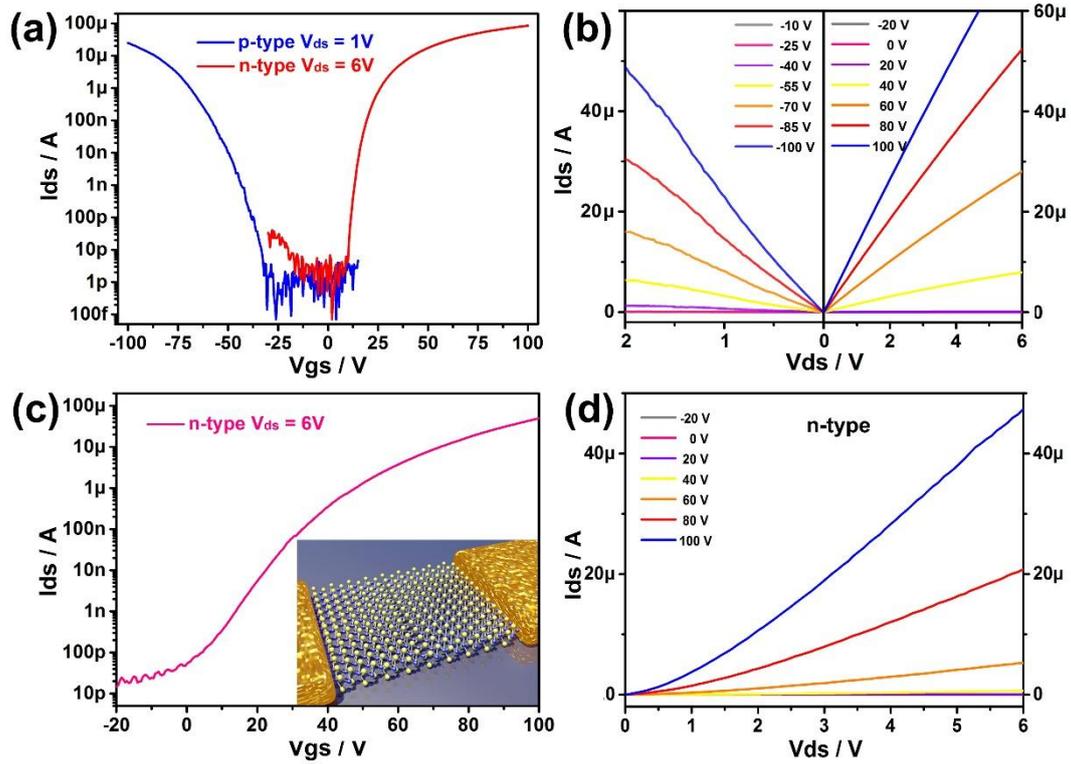

**Figure 4.** Transfer characteristics of (a) p-type (blue) and n-type (red) WSe$_2$ FETs. P-type and n-type devices were prepared using Pd/Au and Ag/Al/Au as contacts, respectively. (b) Corresponding output curves of the WSe$_2$ devices: p-type (left panel) and n-type (right panel). (c) Transfer and (d) output characteristics of n-type WS$_2$ FET with Cr/Au contacts. Inset of (c) shows a schematic illustration of a WS$_2$ FET on Si/SiO$_2$ substrate. Detailed device dimensions are shown in Figure S5.

# Supplementary Material

## Halide-Assisted Atmospheric Pressure Growth of Large $WSe_2$ and $WS_2$ Monolayer Crystals

Shisheng Li, Shunfeng Wang, Dai-Ming Tang, Weijie Zhao, Huilong Xu, Leiqiang Chu, Yoshio Bando, Dmitri Golberg and Goki Eda

Table S1. Summary of recent advances in growing atomically-thin $WSe_2$ flakes

| Year & Reference | Method, Precursors & Substrates | Furnace Temp. Pressure, Substrate Temp. Carrier gas | Shape, layer & domain size | FET device current on/off ratio & mobility |
|---|---|---|---|---|
| 2013 [1] | VT $WSe_2$ Sapphire | 940 °C, A.P. 6 cm from center Ar | Triangle Mono-, bi-, tri-, multi-layer, ~ 5 μm | N.A. |
| 2013 [2] | CVD $WO_3$ + Se (powder) Sapphire | 925 °C, 1 Torr 750-850 °C Ar+$H_2$ | Triangle Mono-, bi-layer, ~50 μm | Ion-gel gated Monolayer device: ~$10^5$, h: 90 $cm^2$/Vs, e: 7 $cm^2$/Vs |
| 2014 [3] | VT $WSe_2$ (powder) Si/$SiO_2$ | 940 °C, 5 Torr 750-850 °C Ar+$H_2$ | Triangle Monolayer ~30 μm | N.A. |
| 2014 [4] | CVD $WO_3$ + Se + S (powder) Si/$SiO_2$ | 875-925 °C, A.P. 875-925 °C Ar+$H_2$ | Triangle, hexagonal Mono-, bi-, and multi-layer, <10 μm | Back-gated multi-layer device ~$10^6$, ~44 $cm^2$/Vs |
| 2015 [5] | CVD $W(CO)_6$ + $(CH_3)_2Se$ Graphene, Si/$SiO_2$, Sapphire & BN | 600-900 °C, 100-700 Torr 600-900 °C $N_2$+$H_2$ | Triangle Mono-, few-layer ~8 μm | N. A. |
| 2015 [6] | CVD $WO_3$ + Se (powder) Si/$SiO_2$ | 850-1050 °C, A.P. 850-950 °C Ar+$H_2$ | Triangle & hexagonal Mono-, few-layer Monolayer ~20 μm Few-layer ~40 μm | Back-gated monolayer device ~$10^5$, h & e: 1~10 $cm^2$/Vs |
| 2015 [7] | VT $WSe_2$ Si/$SiO_2$ | 1060 °C, A.P. 750-795 °C Ar | Triangle & hexagonal Mono-,bi-, few-layer Monolayer ~10 μm Bilayer ~20 μm Few-layer ~50 μm | Back-gated monolayer device ~$10^8$, h: 100 $cm^2$/Vs few-layer device ~$10^8$, h: 350 $cm^2$/Vs |
| 2015 [8] | CVD $WO_3$ + Se (powder) Sapphire, Si/$SiO_2$ | 900-950 °C, A.P. 900-950 °C Ar+$H_2$ | Triangle, hexagonal Mono-, multi-layer <20 μm | N.A. |
| $WSe_2$ Our work | CVD $WO_{2.9}$ + halides + Se (powder) Sapphire, Si/$SiO_2$ | 700-850 °C, A.P. 700-850 °C Ar+$H_2$ | Triangle, hexagonal Monolayer Triangle ~50 μm Hexagonal ~140 μm | Back-gated monolayer device >$10^7$ h: 102 $cm^2$/Vs e: 26 $cm^2$/Vs |

Table S2. Summary of recent advances in growing atomically-thin WS$_2$ flakes

| Year & Reference | Method, Precursors & Substrates | Furnace Temp. Pressure Substrate Temp. Carrier gas | Shape, layer & domain size | FET device current on/off ratio & mobility |
|---|---|---|---|---|
| 2013 [9] | CVD<br>WO$_3$ + S (powder)<br>Si/SiO$_2$ | 800 °C, A.P.<br>800 °C<br>Ar. | Triangle<br>Mono-, bi- and tri-layer<br><20 μm | N.A. |
| 2013 [10] | CVD<br>WO$_3$ + S<br>WO$_3$ film deposits on Si/SiO$_2$ | 800 °C, A.P.<br>800 °C<br>Ar. | Triangle<br>Mono-, bi-, tri- and multi-layer<br><10 μm | N.A. |
| 2013 [11] | CVD<br>WO$_3$ + S<br>Sapphire | 900 °C, 30 Pa<br>900 °C<br>Ar/H$_2$ | Triangle<br>Mono-, bi-, tri-layer<br><50 μm | Ion-gel gated monolayer device:<br>~10$^2$, h: 0.28 cm$^2$/Vs,<br>e: 0.46 cm$^2$/Vs |
| 2013 [12] | CVD<br>WO$_3$ + S (PTAS seeds)<br>Si/SiO$_2$ | 800 °C, A.P.<br>800 °C<br>Ar | Triangle<br>Mon-, Few-layer<br>< 20 μm | Back-gated monolayer device<br>~10$^5$, e: ~0.01 cm$^2$/Vs |
| 2014 [13] | CVD<br>WO$_3$ + S (powder)<br>WO$_3$ powder on Si/SiO$_2$ | 750 °C, A.P.<br>750 °C<br>Ar | Triangle<br>Monolayer<br>~180 μm | N.A. |
| 2014 [14] | CVD<br>WCl$_6$ + S (powder)<br>hBN | 900 °C, A.P.<br>900 °C | Triangle<br>Monolayer<br><5μm | N.A. |
| 2014 [15] | CVD<br>WO$_3$ + S (powder)<br>Si/SiO$_2$ | 1070 °C, A.P.<br>~860 °C<br>Ar. | Triangle<br>Monolayer<br>~370 μm | N.A. |
| 2015 [16] | CVD<br>WO$_3$ + S (powder)<br>Sapphire, Si/SiO$_2$ | 900 °C, A.P.<br>18 cm from center<br>Ar+H$_2$ | Triangle<br>Monolayer<br>~135 μm | Back-gated monolayer device<br>~10$^5$, e: ~4 cm$^2$/Vs |
| 2015 [17] | CVD<br>WO$_3$ + S (powder)<br>Si/SiO$_2$ | 850-900 °C, A.P.<br>850-900 °C<br>Ar+H$_2$ | Triangle<br>Monolayer<br>~50 μm | N.A. |
| 2015 [18] | CVD<br>Ammonium metatungstate hydrate + H$_2$S<br>Au | 935 °C, A.P.<br>935 °C<br>H$_2$S | Triangle<br>Monolayer<br>~420 μm | Back-gated monolayer device<br>~10$^8$, ~20 cm$^2$/Vs |
| Our work | CVD<br>WO$_{2.9}$ + halides + S (powder)<br>Si/SiO$_2$ | 700-850 °C, A.P.<br>700-850 °C<br>Ar+H$_2$ | Triangle, hexagonal<br>Monolayer<br>Triangle ~80 μm<br>Hexagonal ~200 μm | Back-gated monolayer device<br>>10$^6$, e: 14.2 cm$^2$/Vs |

VT: vapor transport: CVD: chemical vapor deposition; FET: field effect transistor; Temp.: Temperature; A.P.: atmospheric pressure; N.A.: not available.

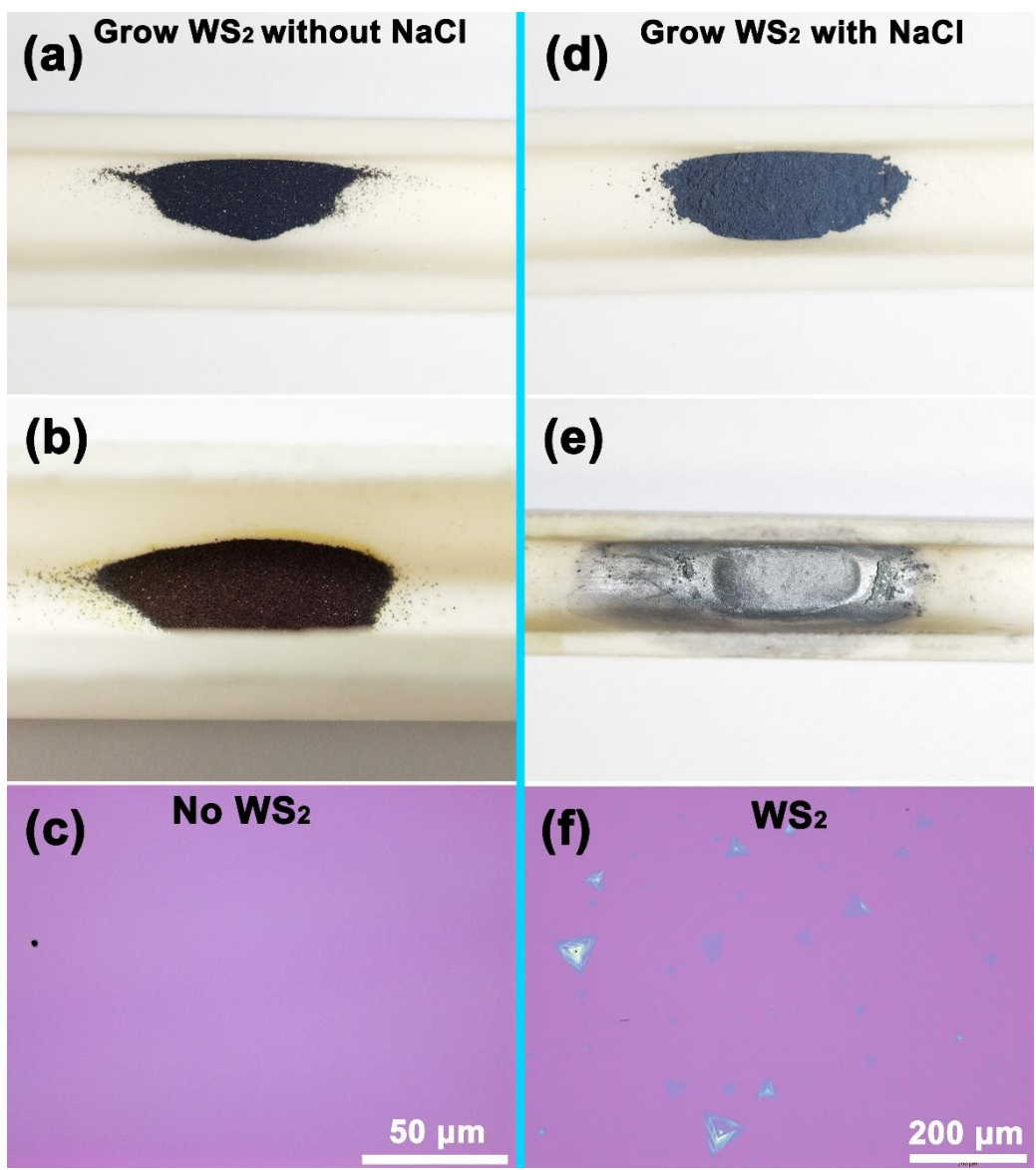

**Figure S1.** Optical microscopy (OM) images of $WO_{2.9}$ powders (a) before and (b) after a typical chemical vapor deposition (CVD) of $WS_2$ process (850 °C, 80 sccm Ar/20 sccm $H_2$). Only a slight color change of $WO_{2.9}$ powders (c) OM image of the $Si/SiO_2$ substrate after $WS_2$ growth, no $WS_2$ flakes deposited. OM images of $WO_{2.9}$/NaCl (70/30 mg) powders (d) before and (e) after a typical CVD of $WS_2$ process (850 °C, 80 sccm Ar/20 sccm $H_2$). Shining sliver color of the mixed powder after the CVD process, indicating sufficient reaction among $WO_{2.9}$, NaCl and S. (f) An OM image of $Si/SiO_2$ substrate after $WS_2$ growth; large, triangular $WS_2$ flakes were grown.

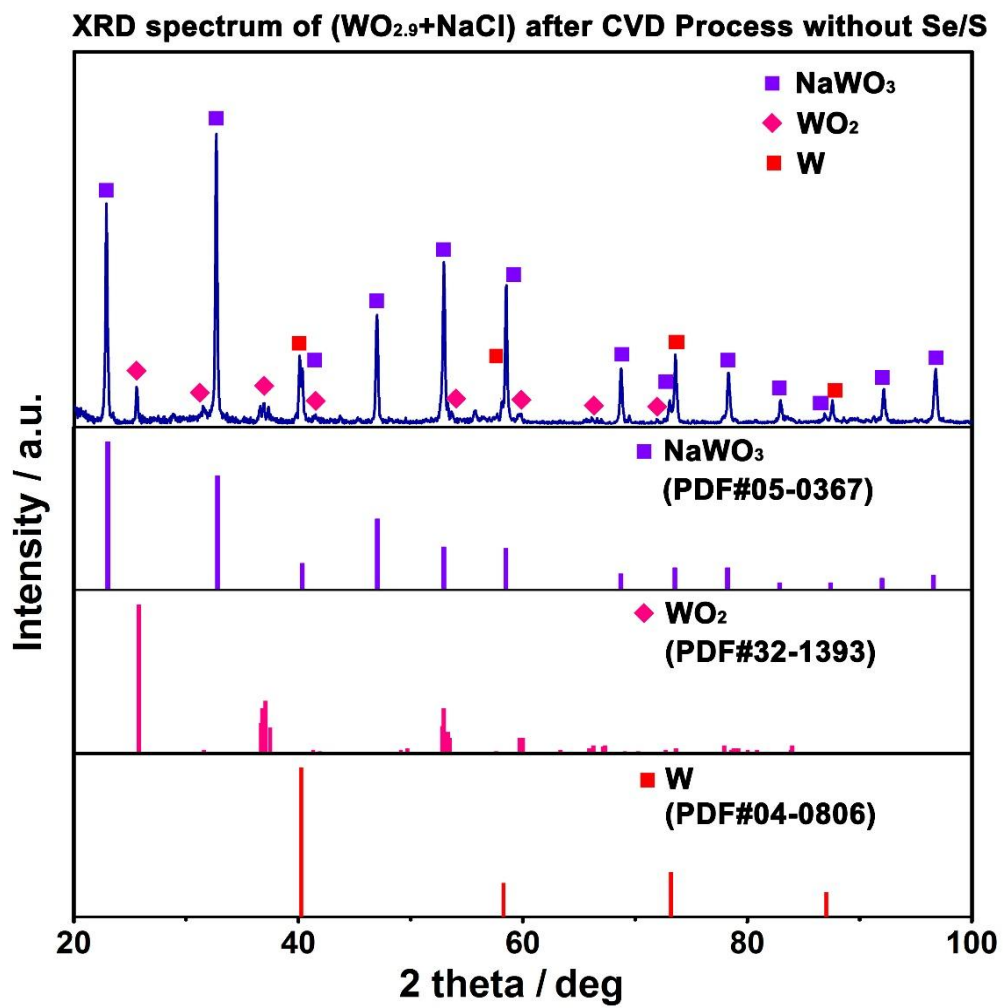

**Figure S2.** X-ray powder diffraction pattern for the mixture of $WO_{2.9}$ and NaCl after a typical CVD process without Se/S at 850 °C under atmospheric pressure. Bottom traces represent the powder diffraction for $NaWO_3$, $WO_2$ and W standard.

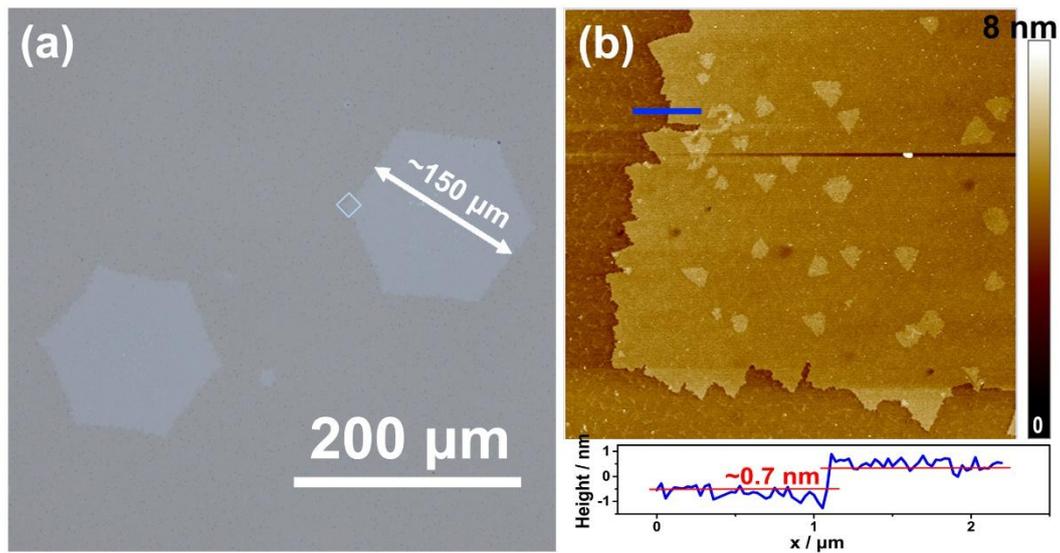

**Figure S3** (a) An OM image of hexagonal WSe$_2$ monolayers grown by NaCl assisted CVD at 850 °C on sapphire substrate. (b) AFM image and height profile of the framed part of theWSe$_2$ monolayer of (a). A mixture of 100 mg WO$_{2.9}$/NaCl powders (85/15 mg) and ~40 mg Se were used.

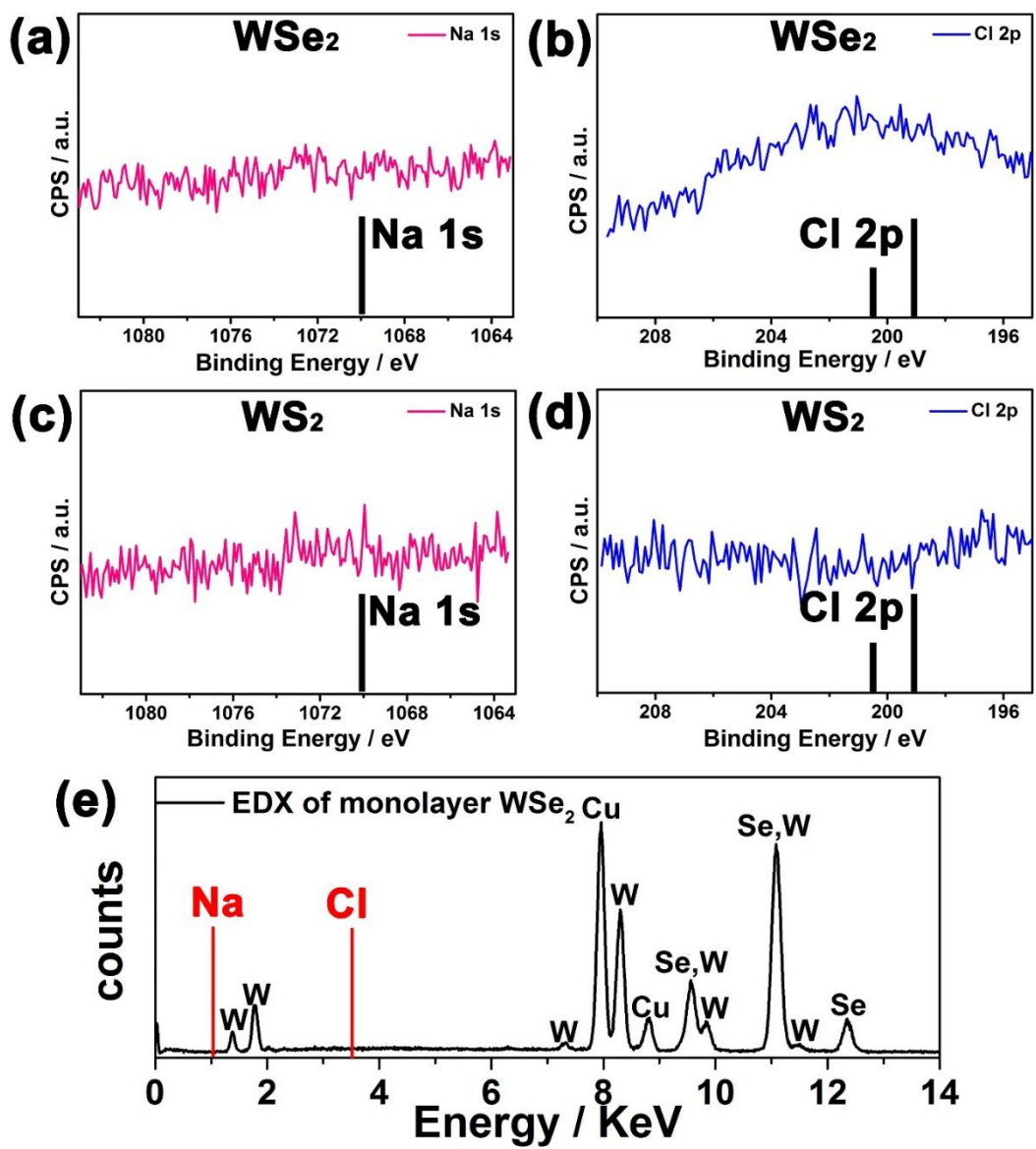

**Figure S4.** The XPS spectra for (a) Na *1s*, (b) Cl *2p* of the as-grown WSe$_2$ monolayers and (c) Na *1s*, (d) Cl *2p* of the as-grown WS$_2$ monolayers. No characteristic peaks of Na *1s* and Cl *2p* were detected. (a) Energy-dispersive X-ray spectroscopy (EDX) data for a WSe$_2$ domain taken in TEM. No Na K line and Cl K line were observed in the spectrum indicating high-purity of the as-grown WSe$_2$ monolayers.

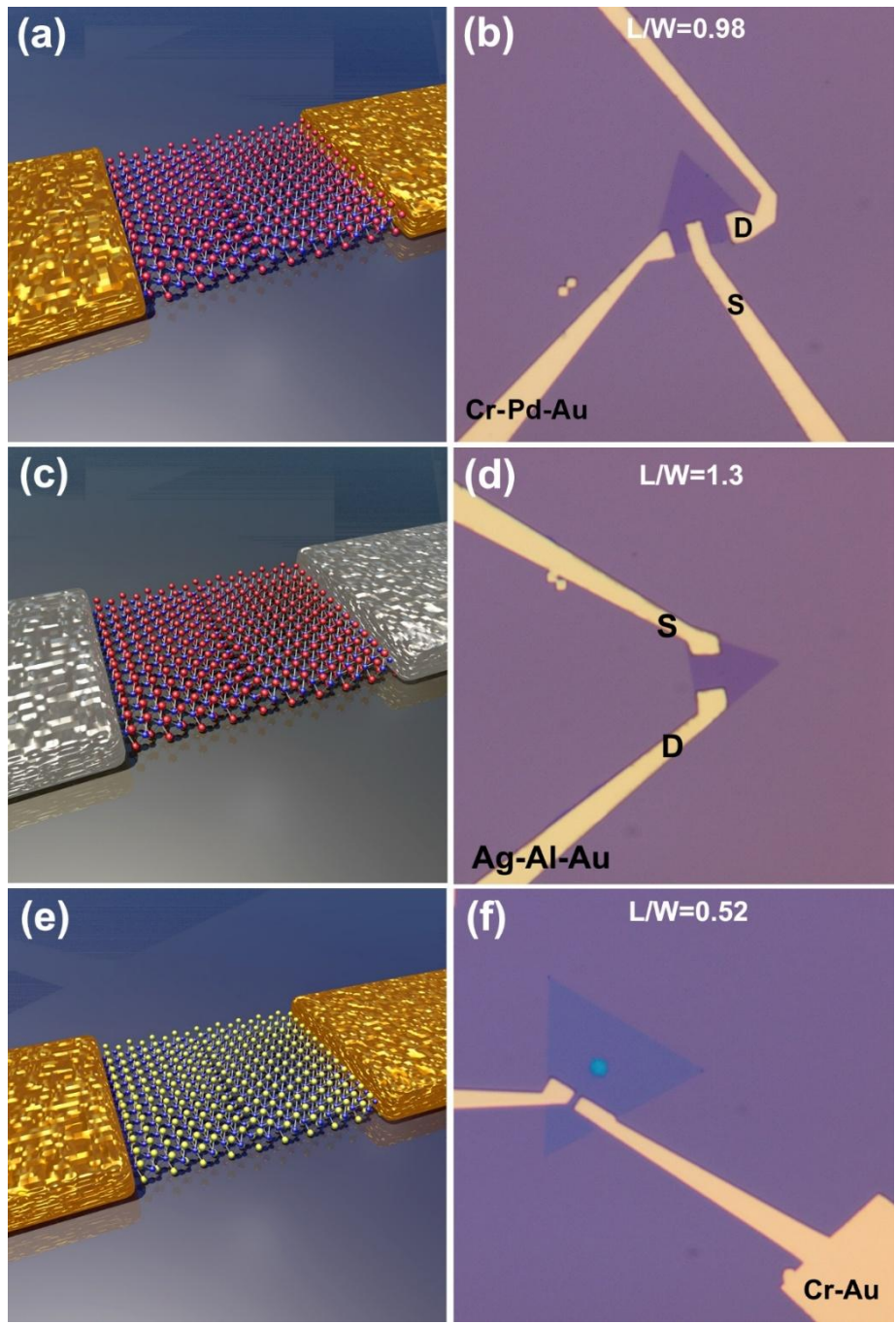

**Figure S5.** Schematic cartoons and corresponding OM images of (a, b) WSe$_2$ p-type devices with Cr/Pd/Au contacts, (c, d) n-type devices with Ag/Al/Au contacts and (e, f) WS$_2$ n-type devices with Cr-Au contacts on Si/SiO$_2$ substrates. The channel length/width for (b) L/W= 2.75 μm/2.8 μm, (d) L/W = 4.2 μm/3.3 μm and (f) L/W= 0.9 μm/1.75 μm.

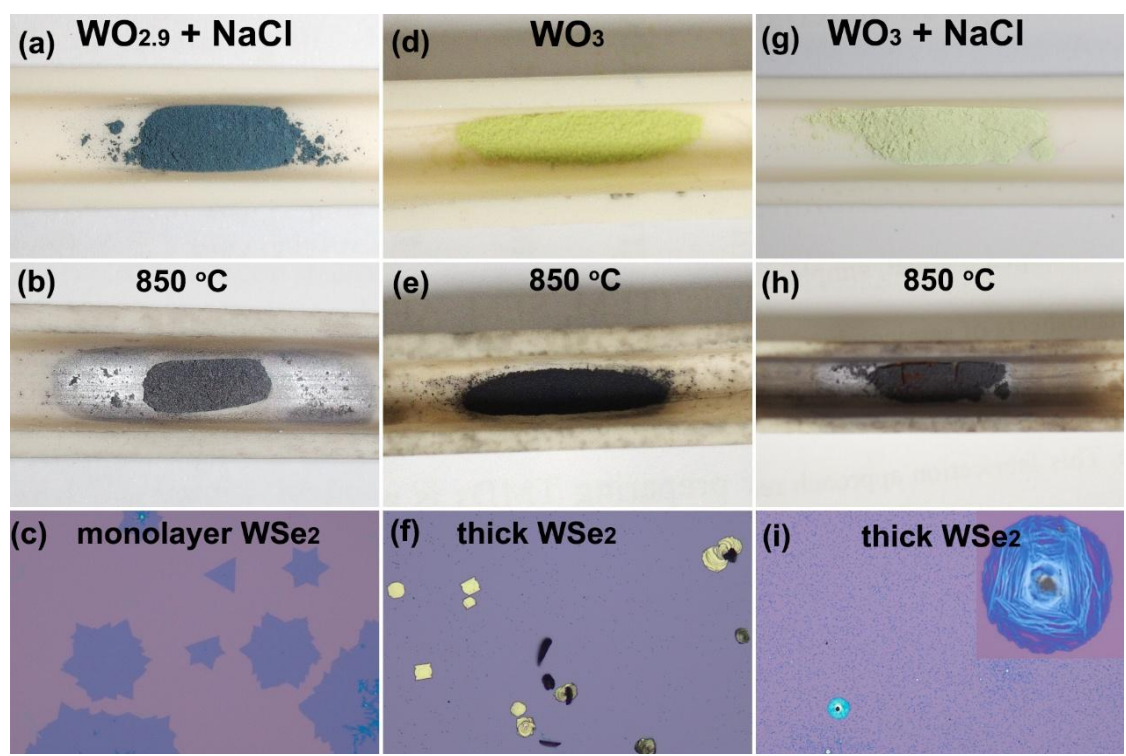

**Figure S6.** (a), (d) and (g) are OM images of WO$_{2.9}$/NaCl, WO$_3$ and WO$_3$/NaCl powders, respectively. (b), (e) and (h) are corresponding OM images of the powders after growing WSe$_2$ at 850 °C, atmospheric pressure. (c), (f) and (i) are corresponding OM images of the WSe$_2$ samples grown from WO$_{2.9}$/NaCl, WO$_3$ and WO$_3$/NaCl, respectively.